# Modeling of Magneto-Conductivity of Bismuth Selenide - A Topological Insulator

Yogesh Kumar[1,2], Rabia Sultana[3], Prince Sharma[1,2,] and V.P.S. Awana[1,2*]

[1]*CSIR-National Physical Laboratory, Dr.K.S. Krishnan Marg, New Delhi-110012, India*
[2]*Academy of Scientific and Innovative Research (AcSIR), Ghaziabad 201002, India*
[3]*Asutosh College, University of Calcutta, Kolkata, India-700026*

**Abstract**

We report the magneto-conductivity analysis of $Bi_2Se_3$ single crystal at different temperatures in a magnetic field range of ±14Tesla. The single crystals are grown by the self-flux method and characterized through X-ray diffraction, Scanning Electron Microscopy, and Raman Spectroscopy. The single crystals show magnetoresistance (MR%) of around 380% at a magnetic field of 14T and a temperature of 5K. The Hikami-Larkin-Nagaoka (HLN) equation has been used to fit the magneto-conductivity (MC) data. However, the HLN fitted curve deviates at higher magnetic fields above 1 Tesla, suggesting that the role of surface driven conductivity suppresses with an increasing magnetic field. This article proposes a speculative model comprising of surface-driven HLN and added quantum diffusive and bulk carriers driven classical terms. The model successfully explains the MC of the $Bi_2Se_3$ single crystal at various temperatures (5 – 200K) and applied magnetic fields (up to 14Tesla).

[*]**Corresponding Author**

Dr. V. P. S. Awana:  E-mail: awana@nplindia.org
Ph. +91-11-45609357, Fax-+91-11-45609310
Homepage: awanavps.webs.com

**Introduction**

Topological Insulators have gained tremendous attention in recent years due to helical spin texture exhibiting vital quantum transport phenomena [1,2]. The presence of strong spin-orbit coupling (SOC) in these materials causes band inversion that results in Dirac-cone-like surface states (SS), which is protected by time-reversal symmetry (TRS) [1-9]. In momentum space, the surface carriers acquire a π Berry phase when moving adiabatically around the Fermi surface. This additional π Berry phase results in destructive interference between two time-reversed paths [10-12]. It suppresses the backscattering of electrons from static scattering centers and enhances the conductivity with decreasing temperature. This electronic transport property in solids is classified by phase coherence length ($L_\varphi$), which depends on temperature. It becomes more extensive than the mean free path (L), i.e., $L_\varphi \gg L$ at low temperatures [13]. With an increase in $L_\varphi$, quantum enhancement is added to classical electronic conductivity, leading to topological delocalization or Weak anti-localization effect (WAL) [14,15]. It is considered as a hallmark for topological surface states (SSs) and is expected to be in systems with simplistic symmetry [16].

Experimentally the SS in topological insulators can be determined by Scanning Tunnelling Microscopy (STM), and Angle-Resolved Photoemission Spectroscopy (ARPES) [17-19]. Numerous bismuth (Bi) based materials such as $Bi_{1-x}Sb_x$, $Bi_2Te_3$, and $Bi_2Se_3$ have been experimentally verified to possess these topological SS [20,21], and their existence in these 3D binaries TIs have been verified through magneto-transport measurements in thin films [22], and single crystals [23]. The surface states of these topological insulators, i.e., $Bi_2Te_3$ and $Bi_2Se_3$, are verified to possess only one helical Dirac cone [4,5,18], and WAL is intrinsic to them. Some of us earlier observed the signatures of the WAL effect in $Bi_2Se_3$ topological insulators by analyzing the magneto-conductivity data using Hikami-Larkin-Nagaoka (HLN) equation [24]. Interestingly, the applicability of HLN was found to be effective only at low applied fields of maximum up to, say 2 Tesla [24].

This article explores the magneto-transport properties of $Bi_2Se_3$ single crystal under an applied magnetic field of up to 14Tesla at different temperatures from 200K down to 5K. The single crystals of $Bi_2Se_3$ were grown by the self-flux method. The material shows magnetoresistance of ~380% at 5K. The magneto-conductivity shows a cusp, which signifies the presence of the WAL effect at a low magnetic field. The HLN is fitted and found to be applicable at a low magnetic field only, (≈1 Tesla) [23,25-27], whereas it deviates from experimental data at a higher field. A quadratic term consisting of quantum and classical contributions is added in conventional HLN [28-30], which accounts for the MC in the entire field range. It is suggested that the contribution of bulk carrier starts to dominate above around 100K. However, we find that these quadratic term does not appropriately account for the MC in our case as apart from these quantum scatterings, there is a possible role of bulk carriers that comes into the picture at higher temperatures and applied magnetic fields. Therefore, the MC data have been fitted by adding a linear term in the HLN equation to account for the bulk contribution. With this finding, we discuss the vital role of bulk conductivity contributions. Summarily, it is shown that two-dimensional transport features may not always originate from the surface states alone in Bi-based topological insulators.

**Experimental Details**

The $Bi_2Se_3$ single crystal was grown using a solid-state reaction by the self-flux method. The high purity (99.99%) bismuth (Bi) and selenium (Se) powders from Alfa Aesar, USA, were taken in the stoichiometric ratio. The powders were mixed thoroughly in the inert (Argon) atmosphere and pelletized into a rectangular pellet with a high vacuum of $10^{-5}$ mbar using a rotary and diffusion pump system. It was kept in an auto programmable furnace with optimized heat treatment, described in our previous report [31]. The obtained crystal was mechanically cleaved for further characterization measurements. The X-ray diffraction (XRD) was performed at room temperature using a Rigaku MiniFlex-II X-ray diffractometer with Cu-K$_\alpha$ radiation ($\lambda$ = 1.5418 Å). The SEM (Scanning Electron Microscopy) images were taken with ZEISS EVO MA10, while the Raman modes were confirmed from the Renshaw Raman spectrometer. The magneto-transport measurements were performed on Quantum Design Physical Property Measurement System (PPMS-14Tesla down to 5K) using the standard four-probe method.

**Results and Discussion**

Figure 1 shows the recorded X-ray diffraction pattern of $Bi_2Se_3$ single-crystal flakes. The XRD spectra were recorded for a 2θ range from $10^0$ to $80^0$. The recorded pattern confirms the unidirectional growth of $Bi_2Se_3$ crystals along the (00$l$) plane. The single crystal has a rhombohedral crystal structure and having $R\bar{3}m$ space group, as described in our previous report [31]. The insets of Fig.1 show the Raman spectra and the SEM image of the single crystal. The Raman modes are in accordance with the previously described crystal as the single crystal have three active Raman modes at 72, 131, and 177 cm$^{-1}$, corresponding to $A_{1g}^1$, $E_g^2$ and $A_{1g}^2$ respectively [31]. The SEM image confirms the layered structure of the topological single crystal.

The transport behaviour of Bi$_2$Se$_3$ single crystal is studied by performing the temperature-dependent magnetoresistance (MR) measurements in the perpendicular magnetic field, and the results are shown in Figure.2. The MR (%) is calculated by using the equation

$$MR\ (\%) = \frac{\rho(H) - \rho(0)}{\rho(0)} \times 100$$

where ρ(0) and ρ(H) are the resistivity at zero and applied magnetic fields, respectively. The linear magnetoresistance (LMR) is nearly 380% at 5K, which eventually decreased to 60% as the temperature is increased to 200K.

The giant MR is also observed in other doped single crystals of TI, such as Bi$_{1.1}$Sb$_{0.9}$Te$_2$S, and its doping with Sn and V [32]. These thick crystals show quatum Shubnikov-de Haas oscillations (SdH) even at 50 K that confirm Fermi surface and confirm the 2D like behaviour in these thick samples [32]. It supports the argument of having surface contribution even in thick samples. Comparing the MR of a single crystal with nanosheets and thin films, a similar trend of high magneto-resistance in Bi$_2$Te$_3$ nanosheets is reported by X. Wang *et al.* [33]. A nanosheet of 20 nm shows huge MR even at room temperature, where this linear MR is accounted for quantum magneto-resistance model [33]. Apart from nanosheets, X.Yu *et al.* report the magneto-conductivity behaviour with applied magnetic field and temperature in Bi$_2$Te$_3$ thin films [34]. At the same time, M. Lang *et al.* reported with Bi$_2$Se$_3$ thin films [35]. The 2D metallic SS is observed in these passivated films through SdH oscillations, and the WAL effect is observed in MR. Comparing to the single crystal in our case, there are no such SdH oscillations while the MR confirms the WAL phenomenon.

The single crystal has a v-type cusp at low temperatures in the low magnetic field regime, which corresponds to weak anti-localization (WAL), as observed in various other topological materials [22-24,31]. At higher magnetic fields, the MR is non-saturating and varies linearly with the magnetic field. Further, with an increase in temperature (5K to 200K), the MR shows a distinctive behaviour with quadratic dependence on low magnetic fields. Correspondingly, at the magnetic field of 14 Tesla, the linear magnetoresistance (LMR) is nearly 380% at 5K, which eventually decreased to 60% as the temperature is increased to 200K. It is because of the reduction of the phase coherence length with an increase in temperature. In previous reports, the LMR is explained classically by Parish and Littlewood (PL) model for disordered systems. According to this model, the fluctuations in carrier mobility will lead to LMR, and also, the crossover field is inversely proportional to carrier mobility [36-39]. However, in the present work, the studied $Bi_2Se_3$ single crystals have layered structures which confirms from SEM images. Therefore the observed MR cannot be explained by the classical PL model.

Further, the magneto-conductivity of the crystal is analyzed to study the role of surface and bulk carriers at different temperatures and magnetic fields. The magneto-conductivity is observed by inversion of resistivity data at their respective fields and temperatures. In magneto-conductivity, a sharp cusp is observed in low magnetic fields and at low temperatures, which broadens with an increase in temperature. Moreover, the magneto-conductivity decreases as the field strength are increased. Figure 3 shows the change in magneto-conductivity (Δσ) in units of $e^2/h$ versus applied perpendicular magnetic field from temperatures 200K down to 5K. Δσ(H) is defined as the difference between magneto-conductivity at applied field [σ(H)] and magneto-conductivity at zero fields [Δσ(0)]. The low field magneto-conductivity is explained by Hikami-Larkin-Nagaoka (HLN) equation in various previous reports [40, 41]. Similarly, the MC data of $Bi_2Se_3$ single crystal at the high perpendicular magnetic field is tried to fit with HLN equation [42], which is given as

$$\Delta\sigma(H) = -\frac{\alpha e^2}{\pi h}\left[ln\left(\frac{B_\varphi}{H}\right) - \Psi\left(\frac{1}{2} + \frac{B_\varphi}{H}\right)\right] \qquad (1)$$

where e is the electronic charge, h is Plank's constant, H is applied magnetic field, Ψ is digamma function, $B_\varphi = \frac{h}{8e\pi L_\varphi^2}$ is characteristic field, and $L_\varphi$ is phase coherence length.

The magneto-conductivity data is fitted by the HLN equation with two fitting parameters, i.e., $L_\varphi$ the phase coherence length and the prefactor α. The negative and positive value of α corresponds to weak anti-localization and weak localization, respectively. In a topological insulator, α = -1/2 corresponds to the single conducting channel and α = -1 indicates two independent conducting channels with each contributing α = -1/2 [43]. Figure 3(a) shows the magneto-conductivity analysis by fitting the HLN equation at all temperatures up to a magnetic field of 1Tesla. A good fit between the experimental data and the HLN equation is obtained in this field region. The obtained value of α is -0.967(7) and -0.810(3) for temperature 5K and 200K, respectively. All the fitted values at different temperatures of prefactor α and phase coherence length $L_\varphi$ are listed in table 1. The value of α ≈ -1 at 5K corresponds to the presence of the WAL effect and contribution in conductivity from two conducting channels.

The increased value of α from -0.967(7) and -0.810(3) with an increase in temperature suggests a weak coupling between bulk states and surface states. The phase coherence length $L_\varphi$ is decreased from 11.077(4) nm to 6.347(6) nm as the temperature is increased from 5K to 200K. These peculiar temperature dependences of $L_\varphi$ can be determined by inelastic electron-phonon or electron-electron scattering. From Fig. 3(a), it is observed that low field magneto-conductivity is well explained by the HLN equation in terms of prefactor α and phase coherence length $L_\varphi$. However, the HLN fitted curve deviates at a higher field (above 1T) from experimental data because of various other contributions apart from surface states.

Hence, we need to establish a precise model to understand the magneto-conductivity of topological insulators in the entire field and temperature ranges. So that in order to have a shred of conclusive evidence regarding the quantum correction in magneto-conductivity. There is a high need to modify the previous MC model. Besides HLN at low field, some additional terms could explain the MC behaviour of topological insulators at higher magnetic fields. In solids, the electronic transport is mainly governed by the phase coherence length $L_\varphi$, spin-orbit scattering length ($L_{SO}$), and mean free path (L). As established from HLN, the phase-coherent term dominates for the low magnetic fields. However, with increasing temperatures, the phase coherence length decreases [44], and the spin-orbit scattering and the elastic scattering term start dominating to conduction. In order to understand the conduction mechanism in the entire magnetic field range, an additional term that is proportional to $H^2$ is added with the HLN equation. This quadratic term eventually corresponds to the elastic/spin-orbit contribution to the conduction channel. Thus, the modified model is given as

$$\Delta\sigma(H) = -\frac{\alpha e^2}{\pi h}\left[\ln\left(\frac{B_\varphi}{H}\right) - \Psi\left(\frac{1}{2} + \frac{B_\varphi}{H}\right)\right] + \beta H^2 \qquad (2)$$

Where the first term is the HLN equation, as explained earlier. The β is a coefficient of the quadratic term that resembles the elastic scattering and spin-orbit scattering in the topological insulator at higher magnetic fields [45]. In addition to these quantum scattering terms, β also comprises of the classical cyclotronic magnetoresistance contribution.

The magneto-conductivity is fitted with this proposed model for the entire field range of magnetic fields of up to 14 Tesla at all temperatures. Figure 3(b) shows the fitted curves of $Bi_2Se_3$ single crystal from temperature 200K down to 5K. The fitted curve resembles the experimental data for the entire field range. It is seen that the HLN fitted curve starts to deviate from experimental data around 1Tesla, and also, the linear magnetoresistance is seen to arise at high field. By using the quadratic term with the HLN equation, the data is well fitted for the entire field range. The fitting parameters α, β, and $L_\varphi$, at different temperatures, are listed in Table 2. The value of α is -0.659(6) at 5K, suggesting the WAL effect. While, with an increase in temperature from 5K to 200K, there is an increase in the α value. The α becomes -0.426(1) at 200K. The positive values of β correspond to the progressive corrections to the HLN equation. It is observed that with increasing temperature from 5K to 200K, there is a slight change in α value (≈ 0.2), and the obtained values of β are of the order of $10^{-4}$.

The peculiar behaviour of the proposed model suggests that the addition of the quadratic term in the HLN equation fits the experimental data well. Also, there is a weak contribution (order of $10^{-4}$) from these quantum scatterings, i.e., spin-orbit scattering and elastic scattering in the case of $Bi_2Se_3$. It is well known that quantum phenomena are suppressed by applying a large magnetic field. Apart from these quantum scatterings, there is a massive chance of the existence of bulk carriers that comes into the picture at higher temperatures and applied magnetic fields. The obtained value of α deviates from -1 with an increase in temperature, which also suggests the coupling between surface and bulk states. In topological insulators, the surface carriers and bulk carriers contribute to overall conductivity. Surface charge carriers contribute at low temperatures and low field, whereas the dominance of bulk charge carrier comes in to picture with increasing field and temperature. It is reported that the bulk carriers are expected to contribute to overall conductivity at temperatures above 100K [46-48]. Therefore, to probe the similar effect, i.e., bulk carriers contribution, a term linear is added with the conventional HLN equation in respect of quadratic term. So as to know whether the conduction channel is only due to surface and bulk carriers and there is no elastic/spin-orbit contribution, the MC data is fitted through linear term along with HLN. Henceforth, the proposed model used to evaluate the magneto-conductivity data theoretically is given by

$$\Delta\sigma(H) = -\frac{\alpha e^2}{\pi h}\left[ln\left(\frac{B_\varphi}{H}\right) - \Psi\left(\frac{1}{2} + \frac{B_\varphi}{H}\right)\right] + \gamma H \qquad (3)$$

Here, the first term is the HLN equation, and γ is the linear term coefficient, which governs the bulk contribution in overall conductivity [49]. The above equation is used to fit the magneto-conductivity data at respective temperatures in the entire field range of up to 14 Tesla. The fitted curve precisely overlaps the experimental data, which is shown in Fig. 3(c). The obtained values of fitted parameters α, $L_\varphi$, and γ are listed in Table 3. The value of prefactor α is -0.835(3) at 5K, and the same increases to -0.169(1) at 200K. At 5K, the value of α obtained from alone HLN and HLN + γH fitting suggests the presence of WAL. The obtained values of γ are of the order of $10^{-3}$. The massive change in α value with increasing temperature suggests that the bulk contribution to overall conductivity increases with rising temperatures.

It is clear from the above analysis that there is an effect of scattering and bulk contributions in magneto-conductivity apart from surface driven assistances. So, it is also a possibility that the crystal does not have any surface contributions to MR%. The experimental data are fitted with the quadratic and linear terms alone to study the dependence of MC on these terms in studied fields and temperature ranges. The fitted curves are shown in Fig. 3(d), and the fitting parameters are mentioned in table.4. At a higher field (> 5 Tesla), the fitted curves are in accordance with experimental data. However, the same deviates at lower magnetic fields. It clearly suggests that surface states dominate in the low field conductivity, and their contribution decreases with increasing field strength. It also endorses that surface states drive the MC at low fields and temperatures.

It is confirmed that there is existence of surface states but the adequate fitting with quadratic and linear term confuses the MC modeling. Basically, the basic HLN fitting confirms the existence of surface terms, while the independent analysis by eq.2 and 3 have confirmed the existence of elastic/spin-orbit and bulk contribution. The linear terms and HLN confirm the existence of bulk carriers and surface states in the system, respectively. In contrast, the quadratic term and HLN also assure the elastic/spin-orbit contribution and surface states. However, there is reasonable fitting in both the cases, i.e., HLN+quadrtic term and HLN+linear term. It arose a dilemma as the possibility of having bulk contribution as well as elastic/spin-orbit contribution in the MC of TIs increases. Basically, to investigate this, the physical significance of MC is thoroughly explored by dividing the MC into two different regions, such as a region of surface states that lies from low field to 1 Tesla only and another one is the elastic/spin-orbit scattering and classical contribution states which lies from 1 Tesla to high fields. Therefore, we fit this MC data independently in their respective region. The HLN precisely governs the surface state contribution. However, if we only consider the quadratic term or linear term, these terms do not fit the experimental data independently. Conversely, if we consider a joint contribution from these two terms, mathematically and physically, these MC data fit adequately and give a precise understanding of such a vast MC. The table.5 explains these dilemmas as the unphysical value of $R^2$, such as the negative value, describes zero possibility of fitting the experimental data with independent consideration of only one contribution. Therefore, there is a high need to reconsider the previous description of MC only with the quadratic term. The combined model of HLN, elastic/spin-orbit, and bulk contribution explain the dilemma of MC and eventually give a physical significance along with mathematical feasibility. The Fig. 4 shows the goodness of fit at 5 K plot in which the HLN in a range 0-1 Tesla, while the quadratic and linear terms in a range 1-14 Tesla are fitted independently. However, the HLN fit maximizes at 1 Tesla; therefore, the combined term needs to be addressed only after 1 Tesla.

Finally, the combined effect of quantum contribution, classical contribution, and bulk carriers in overall conductivity is studied [49]. The experimental data are fitted with a combination of all the three physical contributions in the MC of topological insulators. The magneto-conductivity data is fitted by using the comprehensive model, which probe out all the essential characteristics of the topological insulator, and the clubbed equation is given as

$$\Delta\sigma(H) = -\frac{\alpha e^2}{\pi h}\left[ln\left(\frac{B_\varphi}{H}\right) - \Psi\left(\frac{1}{2} + \frac{B_\varphi}{H}\right)\right] + \beta H^2 + \gamma H \qquad (4)$$

The above equation is fitted at all temperatures for the entire magnetic field range of up to 14Tesla. The fitted curve is in accordance with experimental data in the full-field range, see Figure. 5. The fitted values of α, $L_\varphi$, β, and γ are listed in Table 6. At 5K, the obtained value of α is -0.834(7) and increased to -0.128(1) at a temperature of 200K. These values of α suggest the presence of the WAL effect in $Bi_2Se_3$ single crystals. The trend of α values obtained by a comprehensive model fitting using a clubbed equation is similar to that obtained by fitting the HLN + γH equation.

It is clear that in $Bi_2Se_3$ single crystals, there is a dominance of bulk contribution over quantum and classical scattering in overall conductivity with increasing temperature and magnetic field. Even at high temperatures (200K) and fields, it is confirmed from the MC modeling that the bulk carriers also contribute to conductivity. The classical correction term (γH) and quadratic term ($\beta H^2$) are added to the conventional HLN model to account for temperatures (5K to 200K) and fields (up to 14Tesla) in order to explain the magneto-conductivity of a bulk single crystalline $Bi_2Se_3$ topological insulator which was missing previously. Through this amended modeling, the MC data is explained in terms of the surface, spin-orbit and as well the bulk contributions. The magneto-conductivity (MC) data is fitted with the HLN equation in a field range of 1 Tesla, which confirms the existence of surface states. Further modified HLN is used to study MC by adding a quadratic term $\beta H^2$, which accounts for the elastic/spin-orbit scattering, and a linear term γH, accounting for bulk contribution. When obtained MC is fitted with $\beta H^2$ and γH term alone, the fitting is not converging, concluding that both elastic/spin-orbit scattering and bulk contribution should be taken simultaneously. Hence, finally, the MC is fitted with a combined equation, i.e., HLN+$\beta H^2$+$\gamma H$. This theoretical amendment explains the respective contributions and thereby indirectly predicts the reason for huge MC.

**Conclusion**

We concluded that magneto-conductivity of $Bi_2Se_3$ single crystals at different temperatures for a magnetic field range of ±14Tesla could be modeled by considering the contributions from surface-driven states, quantum scatterings, and the bulk. At low fields, the cusp like behaviour in magneto-conductivity is well explained by the HLN equation, confirming the existence of surface-driven states. However, at higher fields, the HLN fitted curve deviates from experimental data due to quantum scatterings and bulk carrier contributions. Hence there is a prominent competition between HLN, quantum scattering, and cyclotronic motion ($\beta H^2$) and classical ($\gamma H$) contributions. By adding linear terms, the modeling is in accordance with experimental data and physically signifies the bulk carriers contribution. This article substantially confirms the importance of the linear term in the MC modeling and also signifies the increase of bulk contribution with temperature.

**Acknowledgment**

The authors would like to thank the Director of the National Physical Laboratory (NPL), India, for his keen interest in the present work. The author would like to thank CSIR and UGC, India, for a research fellowship and AcSIR-NPL for Ph.D. registration.

**Table 1:** HLN fitted parameters up to ±1Tesla

| Temperature(K) | α | $L_\varphi$(nm) | $R^2$(Goodness of Fit) |
|---|---|---|---|
| 5 | -0.967(7) | 11.077(4) | 0.9768 |
| 50 | -0.975(2) | 9.426(1) | 0.9153 |
| 100 | -0.907(3) | 7.958(5) | 0.9711 |
| 200 | -0.810(3) | 6.347(6) | 0.8191 |

**Table 2:** HLN + $\beta H^2$ fitted parameters up to ±14Tesla

| Temperature(K) | α | $L_\varphi$(nm) | β | $R^2$(Goodness of Fit) |
|---|---|---|---|---|
| 5 | -0.659(6) | 14.034(8) | $3.779 \times 10^{-4}$ | 0.9991 |
| 50 | -0.638(6) | 12.170(1) | $2.952 \times 10^{-4}$ | 0.9995 |
| 100 | -0.482(5) | 9.650(2) | $1.323 \times 10^{-4}$ | 0.9999 |
| 200 | -0.426(1) | 6.480(7) | $6.283 \times 10^{-5}$ | 0.9998 |

**Table 3:** HLN + γH fitted parameters up to ±14Tesla

| Temperature(K) | α | $L_\varphi$(nm) | γ | $R^2$(Goodness of Fit) |
|---|---|---|---|---|
| 5 | -0.835(3) | 15.597(4) | $1.277 \times 10^{-2}$ | 0.9998 |
| 50 | -0.697(1) | 14.269(9) | $8.922 \times 10^{-3}$ | 0.9999 |
| 100 | -0.370(6) | 11.936(1) | $2.467 \times 10^{-3}$ | 0.9998 |
| 200 | -0.169(1) | 8.391(2) | $1.414 \times 10^{-4}$ | 0.9999 |

**Table 4:** HLN + βH² + γH fitted parameters up to ±14Tesla

| Temperature(K) | α | $L_\varphi$(nm) | β | γ | $R^2$(Goodness of Fit) |
|---|---|---|---|---|---|
| 5 | -0.834(7) | 15.577(8) | $4.701 \times 10^{-6}$ | $1.265 \times 10^{-2}$ | 0.9998 |
| 50 | -0.684(8) | 13.719(8) | $8.861 \times 10^{-5}$ | $6.482 \times 10^{-3}$ | 0.9999 |
| 100 | -0.436(9) | 10.340(3) | $9.334 \times 10^{-5}$ | $7.027 \times 10^{-4}$ | 0.9999 |
| 200 | -0.128(1) | 9.328(4) | $-1.699 \times 10^{-5}$ | $2.513 \times 10^{-4}$ | 0.9998 |

**Table 5.** Fitting parameters were obtained by using different models for varying field ranges at temperature 5K.

| βH² | | | |
|---|---|---|---|
| **Field range (T)** | **β** | **γ** | **$R^2$(Goodness of Fit)** |
| 1-14 | NA | NA | Negative for all the ranges |
| γH | | | |
| 1-14 | NA | NA | Negative value for all the ranges |
| βH² + γH | | | |
| 1-14 | $6.725 \times 10^{-4}$ | $-2.123 \times 10^{-2}$ | 0.9877 |

**Table 6:** βH² + γH fitted parameters up to ±14Tesla

| Temperature(K) | β | γ | $R^2$(Goodness of Fit) |
|---|---|---|---|
| 5 | $6.652 \times 10^{-4}$ | $-2.115 \times 10^{-2}$ | 0.9891 |
| 50 | $3.451 \times 10^{-4}$ | $-1.437 \times 10^{-2}$ | 0.9873 |
| 100 | $1.846 \times 10^{-5}$ | $-5.495 \times 10^{-3}$ | 0.9882 |
| 200 | $-5.035 \times 10^{-5}$ | $-1.047 \times 10^{-3}$ | 0.9928 |

**Figure captions:**

Fig. 1: X-ray diffraction pattern of $Bi_2Se_3$ single-crystal flake acquired using Cu $K_\alpha$ radiation. The diffraction peaks confirm the unidirectional growth of $Bi_2Se_3$ crystals along the (00$l$) plane. The left inset shows the SEM images that confirm the layered structure, and the right inset plots the recorded Raman spectra showing three Raman active modes.

Fig. 2: Magneto-resistance of $Bi_2Se_3$ single crystal measured at different temperatures in a perpendicular magnetic field of up to 14 Tesla. The linear magnetoresistance (LMR) is nearly 380% at 5K, which eventually decreased to 60% as the temperature is increased to 200K at the magnetic field of 14 Tesla.

Fig. 3: Magneto-conductivity analysis of $Bi_2Se_3$ single crystal measured at different temperatures. It shows the change in magneto-conductivity ($\Delta\sigma$) in units of $e^2/h$ versus applied perpendicular magnetic field from temperatures 200K down to 5K. $\Delta\sigma(H)$ is defined as the difference between magneto-conductivity at applied field [$\sigma(H)$] and magneto-conductivity at zero fields [$\Delta\sigma(0)$]. **(a)** HLN fitting up to ±1Tesla. **(b)** HLN + $\beta H^2$ is fitting up to ±14Tesla. **(c)** HLN + $\gamma H$ fitting up to ±14Tesla. **(d)** $\beta H^2 + \gamma H$ fitting up to ±14Tesla.

Fig.4: Magneto-conductivity of $Bi_2Se_3$ single-crystal showing the independent contribution of HLN in a field range of up to ±1Tesla and $\beta H^2 + \gamma H$ term for field range of 1-14 Tesla. It shows the goodness of fit at 5 K plot in which the HLN in a range 0-1 Tesla, while the quadratic and linear terms in a range 1-14 Tesla are fitted independently. However, the HLN fit maximizes at 1 Tesla; therefore, the combined term needs to be addressed only after 1 Tesla.

Fig. 5: The Magneto-conductivity of $Bi_2Se_3$ single-crystal is fitted by HLN + $\beta H^2 + \gamma H$ equation at all temperatures for the entire magnetic field range up to 14 Tesla, and solid curve corresponds to the fitted data. There is a bulk contribution dominance over quantum and classical scattering in overall conductivity with increasing temperature and magnetic field. Even at high temperatures (200K) and fields, it is confirmed from the MC modeling that the bulk carriers also contribute to conductivity. The fitting confirms the addition of classical correction term ($\gamma H$) and quadratic term ($\beta H^2$) to the conventional HLN model to account for temperatures (5K to 200K) and fields (up to 14Tesla).

Figure 1

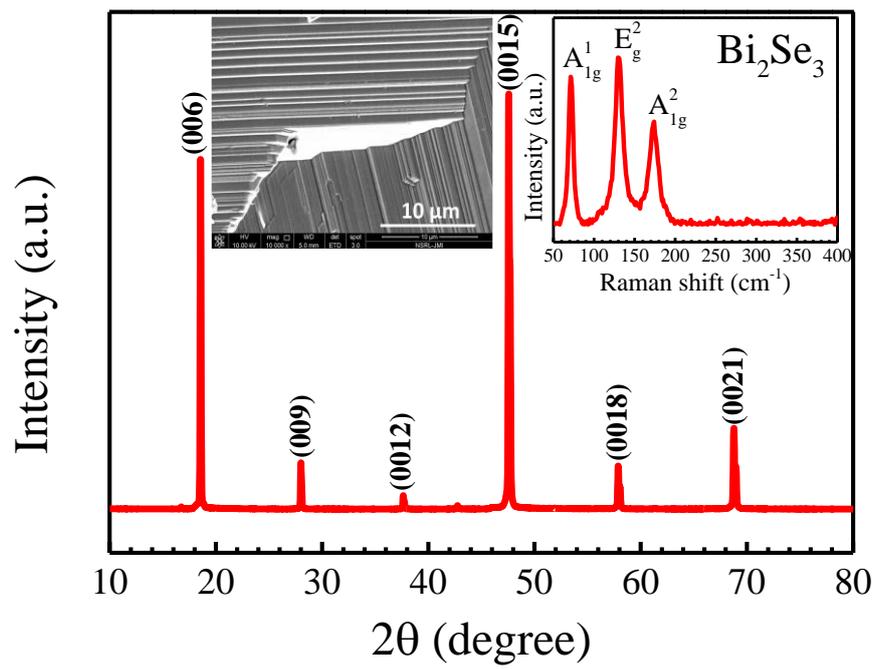

Figure 2

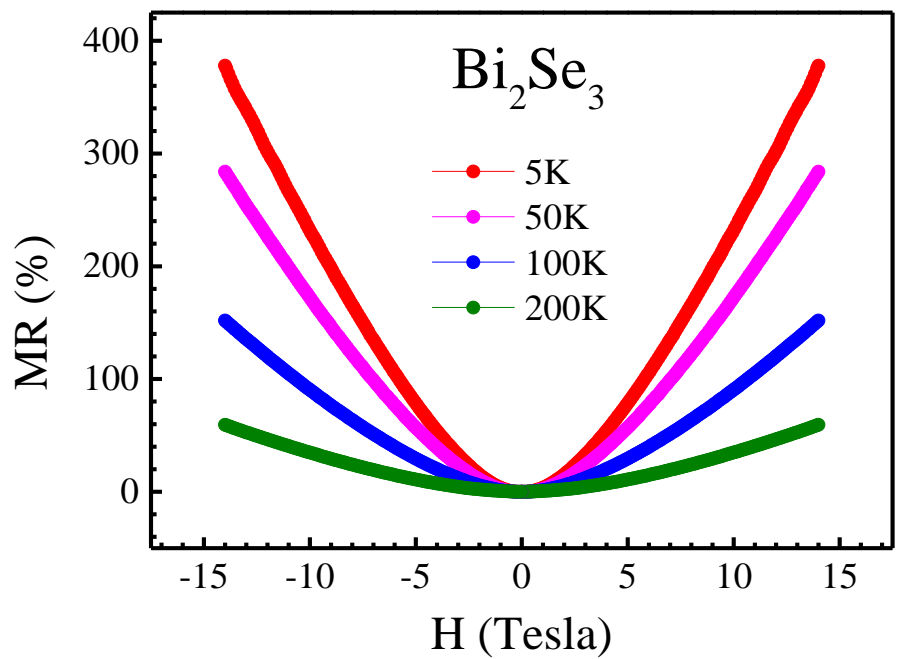

Figure 3 (a)

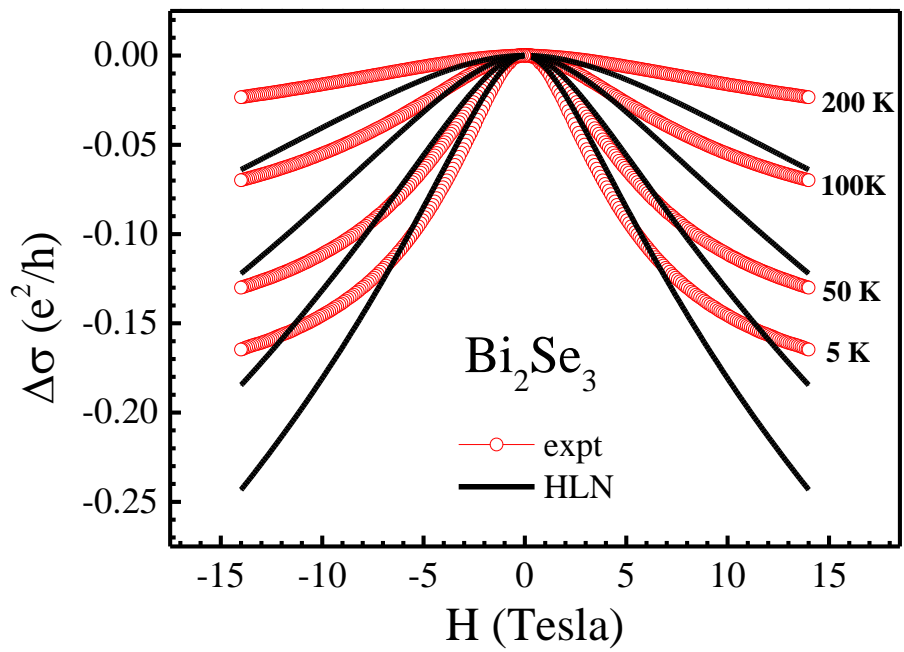

Figure 3 (b)

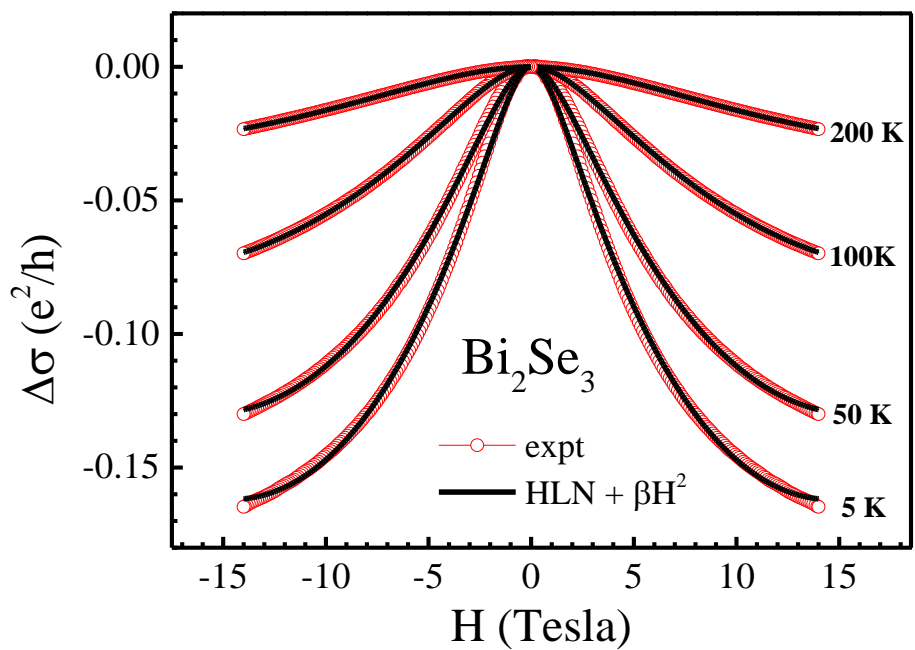

Figure 3 (c)

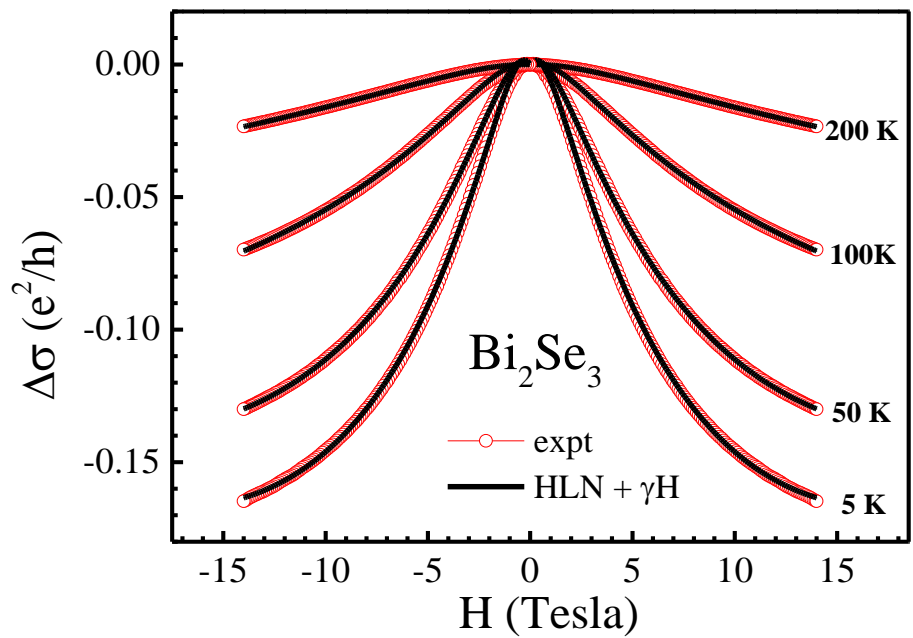

Figure 3 (d)

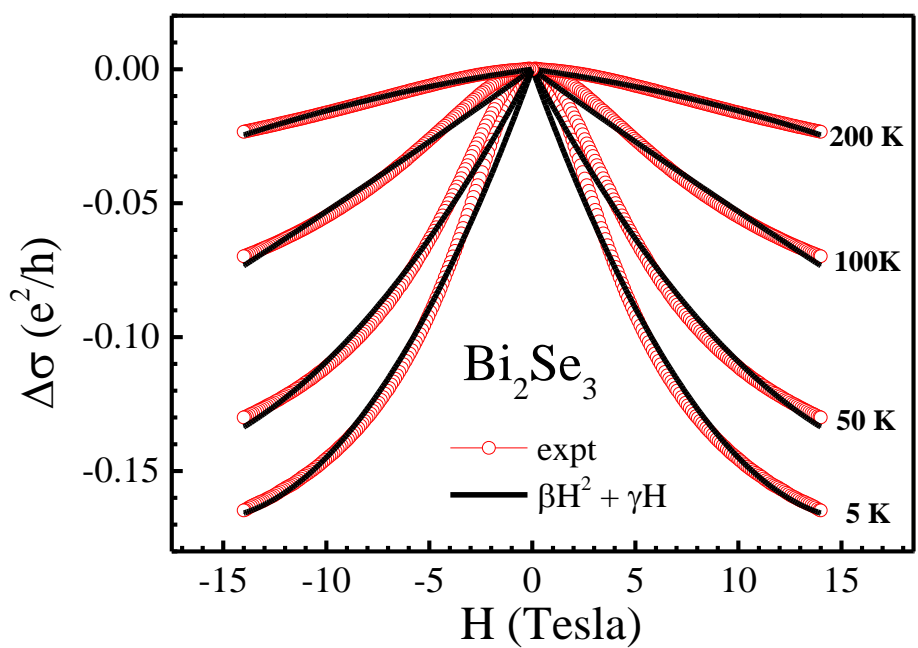

Figure 4

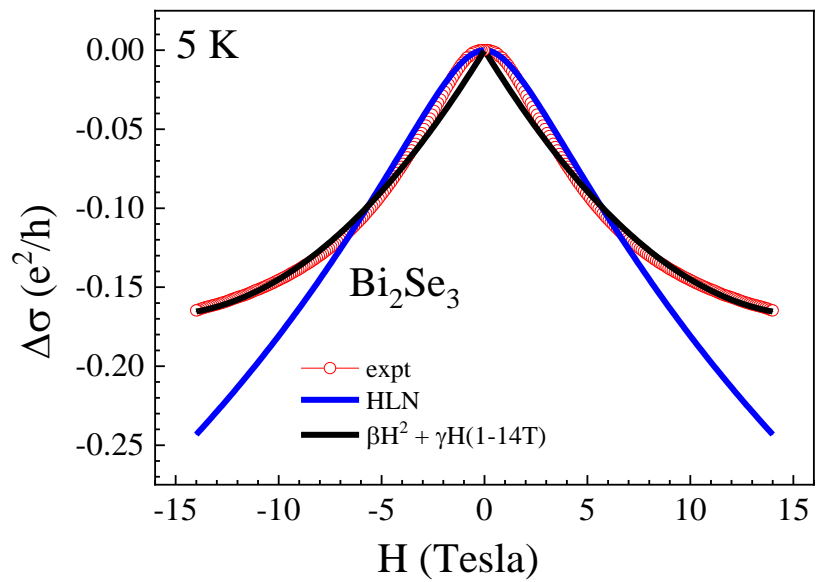

Figure 5

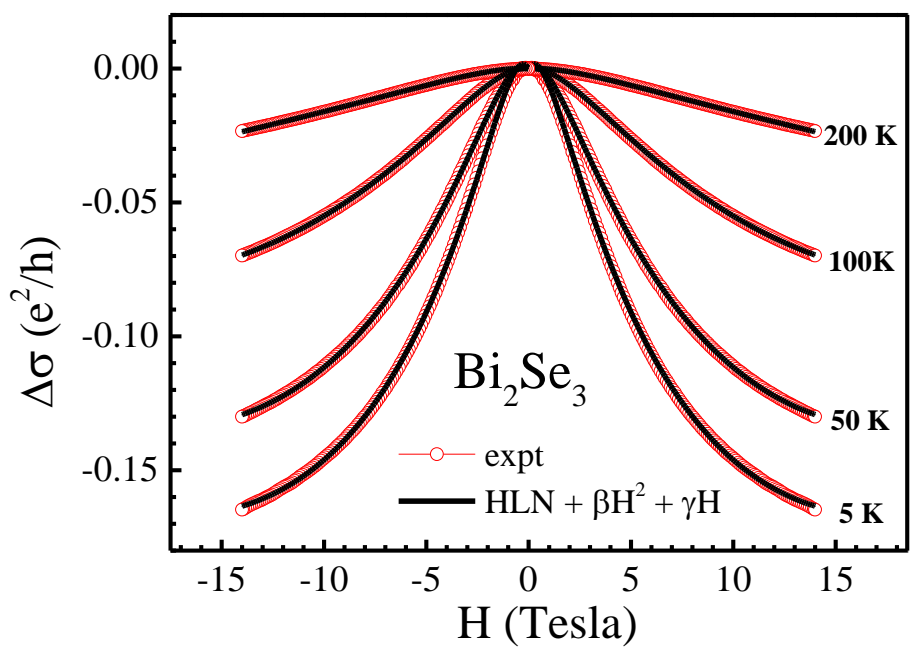